# Infinite Face Centered Cubic Network of Identical Resistors- Application to Lattice Green Function

**J. H. Asad[1], Ashraf Ayed Diab[2], R. S. Hijjawi[3], A. J. Sakaji[4], and J. M. Khalifeh[5]**
[1]Dep. of Physics- Tabuk University- P. O Box 741- Tabuk 71491- Saudi Arabia
[2]General Studies Department- Yanbu Industrial College- P. O Box 30436- Yanbu Industrial City- KSA
[3]Dep. of Physics- Mutah University- Karak- Jordan
[4]Dep. of Physics- Ajman University- UAE
[5]Dep. of Physics-Jordan University- Amman 11942 - Jordan

**Abstract**

The equivalent resistance between the origin and any other lattice site, in an infinite Face Centered Cubic network consisting from identical resistors, has been expressed rationally in terms of the known value $f_o(3;0,0,0)$ and $\pi$. The asymptotic behavior is investigated, and some calculated values for the equivalent resistance are presented.

## I. Introduction

It is well known that the theory of Green function plays an important role in physics. It has been used, e.g., in statistical model of ferromagnetism such as Ising model [1], Heisenberg model [2], spherical model [3,], random walks and percolation theories[4,5], diffusion [6], band structure [7], and other branches in physics. In condensed matter the Lattice Green Function (i.e., LGF) is widely applied and considered a basic function [8]. It appears especially when impure solids are studied [9]. The LGF for several lattices has widely studied and investigated. One can see from the literature that most of the studies on lattice Green functions are based on elliptic type integral approach. In the literature generally, the formulas for the lattice Green functions have been given in terms of elliptic type integrals and related functions.

The LGF for Face Centered Cubic Lattice (i.e., FCC) was studied well by Iwata [10]. He expressed $f_o(3;0,0,0)$ (i.e., the LGF at the origin for a FCC lattice) in a compact form as a product of complete elliptic integrals of the first kind. Much effort has been focused on the value of that function at the origin, although we need also the values at various lattice sites in many problems. Inoue [11] showed that $F(E;l,m,n)$ (i.e., the LGF at an arbitrary lattice site $(l,m,n)$ in FCC lattice) with nearest neighbor interactions can be expressed in terms of linear combinations of products of complete elliptic integrals of the first and second kinds. In his work he introduced some recurrence formula. Morita [12] showed that $F(E;l,m,n)$ can be calculated from the three known values at the lattice sites $(0,0,0), (2,0,0)$, and $(2,2,0)$ with the aid of the recurrence formula presented by Inoue [11]. In a previous work Hijjawi et al [13] present an expression for $F(E;l,m,n)$ where it is evaluated analytically and numerically for a single impurity problem and the density of states, phase shift and scattering cross section are expressed in terms of complete elliptic integrals of the first kind.

The calculation of the effective resistance in infinite networks of identical resistors is a classic problem in the circuit theory. It attracts the attention of many authors since Kirchhoff's [14] who formulated the study of electric networks more than 150 years ago. In addition, the electric circuit theory was discussed in details by Van der Pol and Bremmer [15], where they derived the resistance between nearby points on the square lattice. Many methods have been used to study different infinite networks consisting of identical resistors such as random walk method [16], current distribution method [17-19], and recently, an important method based on the LGF has been introduced by Cserti [20]. Based on this method many studied has been carried out for many networks consisting of identical resistors [21-23]. More recently, one can see that this field is still of interests for many authors where many projects have been carried out by different authors [24,25].

This paper is organized as follows:

In Sect. 2, a short revision is carried out to the origin problem (i.e., an infinite d- dimensional hypercube network consisting of identical resistors). In Sect. 3, we apply the formalism to the FCC network. In Sect. 4, results and conclusions are presented.

## II. Hypercubic Lattices

Consider an infinite d- dimensional lattice consisting of identical resistors each of resistance *R*. We assume that all lattice points in the infinite lattice are specified by the position vector

$$\vec{r} = l_1\vec{a}_1 + l_2\vec{a}_2 + ... + l_d\vec{a}_d .\tag{1}$$

Where $l_1, l_2, ..., l_d$ are integers (positive, negative or zero),

and $\vec{a}_1, \vec{a}_2, ..., \vec{a}_d$ are independent primitive translation vectors.

When all $\vec{a}_i$'s have the same magnitude (i.e. $|\vec{a}_1| = |\vec{a}_2| = ... = |\vec{a}_d| = a$), then the d- dimensional lattice is called a hypercube.

Our aim is to find the equivalent resistance between the origin and any other lattice site $\vec{r}$. To do this, let the potential at the lattice site $\vec{r}'$ be defined as $V(\vec{r}')$, and assume the current in the network defined as:

$$I(\vec{r}') = I[\delta_{\vec{r}',0} - \delta_{\vec{r}',\vec{r}}] .\tag{2}$$

According to Ohm's and Kirchhoff's laws we can write:

$$I(\vec{r}')R_o(\vec{r}') = \sum_n [V(\vec{r}') - V(\vec{r}' + \vec{n})] .\tag{3}$$

Where $\vec{n}$ are the vectors from the site $\vec{r}'$ to its nearest neighbors (i.e., $\vec{n} = \pm\vec{a}_i, i = 1, 2, ..., d$).

Using the so-called lattice Laplacian defined on the hypercubic lattice [ ] as:

$$\Delta_{(\vec{r}')} f(\vec{r}) = \sum_n [f(\vec{r} + \vec{n}) - f(\vec{r})] .\tag{4}$$

One can rewrite Eq. (3) as:

$$\Delta_{(\vec{r}')} V(\vec{r}') = -I(\vec{r}')R_o(\vec{r}') .\tag{5}$$

From Eq. (2) and Eq. (5) we can write:

$$R_o(\vec{r}) = \frac{V(0) - V(\vec{r})}{I} .\tag{6}$$

To find the resistance we need to solve Eq. (5), which is a Poisson-like equation and it may be solved using the LGF, so one may write (comparing with Poisson's equation)

$$V(\vec{r}) = R\sum_{\vec{r}'} G_o(\vec{r} - \vec{r}')I(\vec{r}') .\tag{7}$$

٣

Using Eqs. (2,7) into Eq. (6) we got:
$$R_o(\vec{r}) = 2R[G_o(0) - G_o(\vec{r})]. \tag{8}$$
Where $G_o(\vec{r})$ is the LGF of the d-dimensional hypercube at the lattice site $\vec{r}$, and $G_o(0)$ is the LGF of the d-dimensional hypercube at the origin. The last equation represents the basic result we search. Once, the LGF values are known, one can easily obtain the required equivalent resistance.

Finally, the resistance in a d- dimensional hypercube can be written as [20]:
$$R_o(l_1, l_2, ..., l_d) = R \int_{-\pi}^{\pi} \frac{dx_1}{2\pi} ... \int_{-\pi}^{\pi} \frac{dx_d}{2\pi} \frac{1 - \exp(il_1 x_1 + il_2 x_2 + ... + il_d x_d)}{\sum_{i=1}^{d}(1 - Cosx_i)}. \tag{9}$$

Also, the LGF for a d-dimensional hypercube can be written as [8]:
$$G_o(l_1, l_2, ..., l_d) = \int_{-\pi}^{\pi} \frac{dx_1}{2\pi} ... \int_{-\pi}^{\pi} \frac{dx_d}{2\pi} \frac{\exp(il_1 x_1 + il_2 x_2 + ... + il_d x_d)}{2\sum_{i=1}^{d}(1 - Cosx_i)}. \tag{10}$$

## III- Application: FCC network

Now, we turn to the infinite FCC network which consists of identical resistors each of resistance $R$. In this case the position vector $\vec{r}$ becomes:
$$\vec{r} = l\vec{a}_1 + m\vec{a}_2 + n\vec{a}_3. \tag{11}$$
As a result Eq. (8) can be written as:
$$R_o(l, m, n) = R[f_o(0,0,0) - F_o(l, m, n)]. \tag{12}$$
The LGF for a FCC lattice is defined as:
$$F(E; l, m, n) = \frac{1}{\pi^3} \int_0^{\pi}\int_0^{\pi}\int_0^{\pi} \frac{Cos(lx)Cos(my)Cos(nz)}{E - CosxCosy - CosyCosz - CosxCosz} dxdydz. \tag{13}$$
For $E \geq 3$, and $l + m + n$ even ($F = 0$ for $l + m + n$ is odd) [19,26]

In a recent work [19], the value $F(3; l, m, n)$ has been expressed rationally in terms of the known value $f_o(3;0,0,0)$, and $\pi$ as:
$$F_o(3; l, m, n) = \rho_1 f_o + \frac{\rho_o}{\pi^2 f_o} + \rho_3. \tag{14}$$
Using Eq. (12) and Eq. (14) we can express the equivalent resistance as:
$$\frac{R_o(3; l, m, n)}{R} = r_1 f_o + \frac{r_2}{\pi^2 f_o} + r_3. \tag{15}$$
Where $r_1$, $r_2$ and $r_3$ are rational numbers related to $\rho_1$, $\rho_2$ and $\rho_3$ as:
$$r_1 = 1 - \rho_1, r_2 = -\rho_2 \text{ and } r_3 = \rho_3.$$

and $f_0 = F(3;0,0,0)$ is the LGF of an infinite FCC lattice at the origin. The LGF at the origin for an infinite FCC lattice was first evaluated by Watson where he showed that [27]:

$$f_o = F(3;0,0,0) = \frac{\sqrt{3}}{2}[K(k)]^2 = \frac{3\Gamma^6(\frac{1}{3})}{2^{\frac{14}{3}}\pi^4} = 0.4482203944. \qquad (15)$$

where $k = Sin\frac{\pi}{12} = \frac{\sqrt{3}-1}{2\sqrt{2}}$.

Various values for $r_1$, $r_2$ and $r_3$ can be obtained from Table 2 [19]. Below, we show some of these values in Table 1.

The asymptotic case (i.e., the separation between the origin and the site (l,m,n) goes to a large value or infinity). In this case the resistance goes to a finite value. To explain this point, we note that from the theory of Fourier series (Riemann's Lemma) that $Lim_{n\to\infty}\int_a^b \Phi(x)Cosnx\,dx \to 0$ for any integrable function $\Phi(x)$. Thus, $f_o(l,m,n) \to 0$ (corresponding to the boundary condition of the Green's function at infinity), and as a result Eq. (12) becomes

$$\frac{R_o(l,m,n)}{R} \to f_o. \qquad (16)$$

when any of $l, m, n \to \infty$.

We have calculated additional values for $r_1$, $r_2$ and $r_3$ using the recurrence formulae for the LGF of the infinite FCC lattice [12] for the following lattice sites (6,0,0),(6,2,0),(6,4,0),(7,1,0),(7,3,0), and (8,0,0).

## 3. Results and Discussion

We have expressed the equivalent resistance between the origin and the lattice site (l,m,n) in an infinite FCC network which consists of identical resistors each of resistance $R$. The equivalent resistance is plotted against the lattice site, and from the Figures shown the resistance in an infinite FCC lattice is symmetric under the transformation (l,m,n) → (−l,−m, −n) which is expected due to the inversion symmetry of the lattice.

Figure 1 shows the resistance in an infinite FCC lattice against the site (l, m, n) along the [100] direction. From the figure it is clear that the resistance is symmetric.

°

Figure 2 shows the resistance in an infinite FCC lattice against the site (*l*, *m*, *n*) along the [111] direction. From the figure it is clear that the resistance is symmetric.

From the figures one can see that as the separation between the origin and the lattice site *(l,m,n)* increases then, the equivalent resistance approaches a finite value (i.e., $f_o = 0.4482203944$) as explained above. A similar result was obtained for the resistance in an infinite SC network [22] where as the separation between the origin and any other lattice site the equivalent resistance approaches a finite value (i.e., $g_o = 0.505462$) which is the LGF at the origin in an infinite SC lattice, while the resistance in an infinite square network diverges for large separation between the two sites [21].

# Table Captions

**Table 1:** Values for $r_1$, $r_2$ and $r_3$.

**Table 1:**

| The Site lmn | $r_1$ | $r_2$ | $r_3$ | $R(l,m,n)/R$ |
|---|---|---|---|---|
| 000 | 0 | 0 | 0 | 0 |
| 200 | 4/3 | -1 | 0 | 0.371575 |
| 400 | -16/9 | 16/3 | 0 | 0.408775 |
| 110 | 0 | 0 | 1/3 | 0.333333 |
| 310 | 8/3 | -5 | 1/3 | 0.398327 |
| 510 | -136/9 | 91/3 | 1/3 | 0.41714 |
| 211 | 4/3 | 2 | -2/3 | 0.383065 |
| 411 | 64/9 | -28/3 | -2/3 | 0.410858 |
| 220 | -8 | -6 | 16/3 | 0.391257 |
| 420 | -4/9 | -275/3 | 64/3 | 0.412677 |
| 321 | 28/3 | 29 | -31/3 | 0.405569 |
| 521 | 788/9 | -95/3 | -95/3 | 0.419201 |
| 222 | 4 | -15 | 2 | 0.402099 |
| 422 | -728/9 | 62/3 | 32 | 0.415695 |
| 330 | -144 | -186 | 107 | 0.410563 |
| 530 | -688 | -2316 | 2497/3 | 0.420924 |
| 431 | 2840/9 | 2101/3 | -898/3 | 0.416958 |
| 332 | -4 | -120 | 88/3 | 0.414194 |
| 532 | -8932/9 | -1598/3 | 1697/3 | 0.422381 |
| 433 | 1712/9 | 502/3 | -368/3 | 0.420872 |
| 440 | -9056/3 | -4960 | 7424/3 | 0.42004 |
| 541 | 82936/9 | 56735/3 | -8405 | 0.423639 |
| 442 | -8212/9 | -9059/3 | 1092 | 0.42165 |
| 543 | 159056/45 | 12800/3 | -7645/3 | 0.425684 |
| 444 | -19024/15 | 464 | 464 | 0.425212 |
| 550 | -218480/3 | -133150 | 188225/3 | 0.425693 |
| 552 | -1596212/45 | -249428/3 | 34694 | 0.42654 |
| 554 | -110992/9 | -20534/3 | 21226/3 | 0.428607 |
| 600 | 1924/75 | -49 | 0 | 0.421792 |
| 620 | -56488/225 | 862/3 | 48 | 0.423101 |
| 640 | -2155436/75 | -67681 | 84544/3 | 0.426144 |
| 710 | 1425407/9600 | -294 | 1/3 | 0.425719 |
| 730 | -170742787/28800 | -7201/3 | 9601/3 | 0.427291 |
| 800 | -89212757/352800 | 73984/147 | 0 | 0.428583 |

# Figure Captions

**Fig. 1:** The calculated resistance between the origin and the lattice site (l,m,n) along [100] direction of the infinite FCC lattice

**Fig. 2:** The calculated resistance between the origin and the lattice site (l,m,n) along [111] direction of the infinite FCC lattice

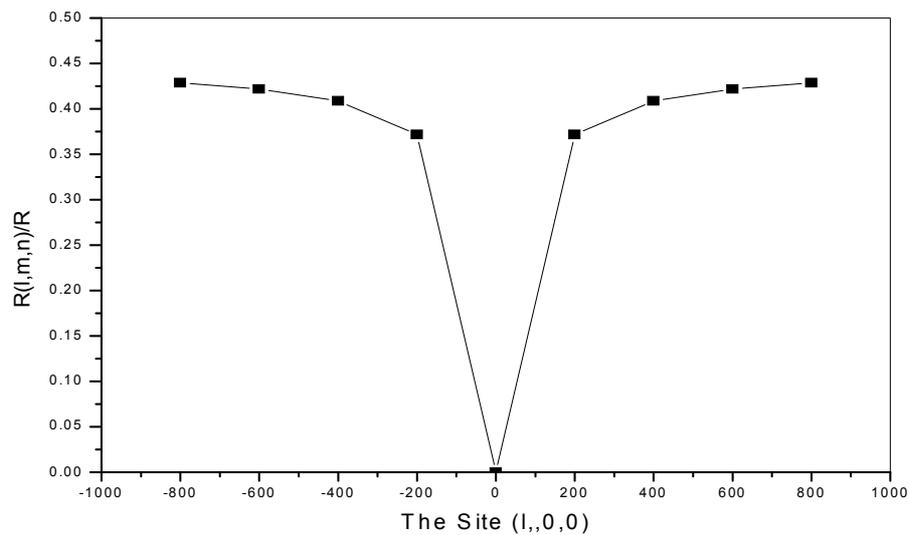

Fig. 1 Resistance between the origin (0,0,0)and the site (l,0,0) along [100]direction for FCC lattice

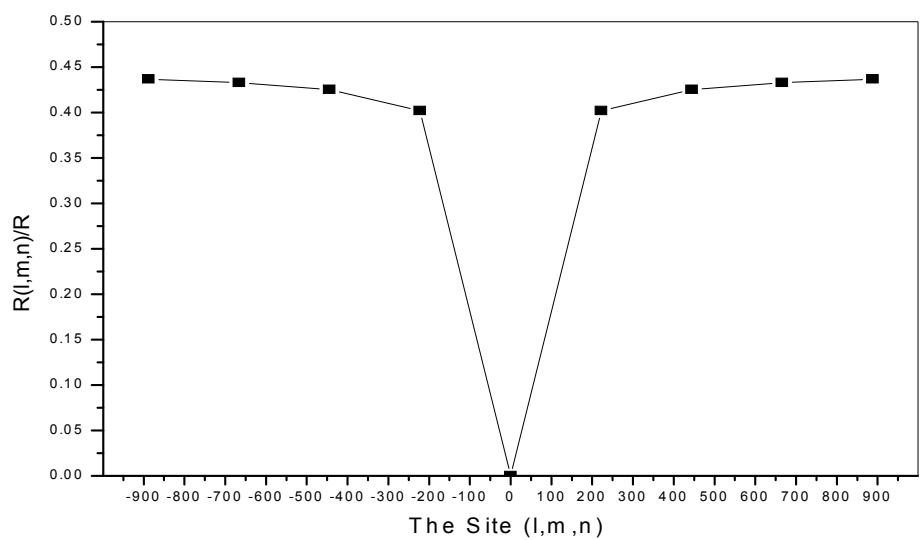

Fig. 2 Resistance between the origin (0,0,0)and the site (l,m,n) along [111]direction for FCC lattice